# SPARK00: A Benchmark Package for the Compiler Evaluation of Irregular/Sparse Codes

H.L.A. van der Spek
E.M. Bakker
H.A.G. Wijshoff
Leiden University, The Netherlands

**Abstract**

We propose a set of benchmarks that specifically targets a major cause of performance degradation in high performance computing platforms: irregular access patterns. These benchmarks are meant to be used to asses the performance of optimizing compilers on codes with a varying degree of irregular access. The irregularity caused by the use of pointers and indirection arrays are a major challenge for optimizing compilers. Codes containing such patterns are notoriously hard to optimize but they have a huge impact on the performance of modern architectures, which are under-utilized when encountering irregular memory accesses. In this paper, a set of benchmarks is described that explicitly measures the performance of kernels containing a variety of different access patterns found in real world applications. By offering a varying degree of complexity, we provide a platform for measuring the effectiveness of transformations. The difference in complexity stems from a difference in traversal patterns, the use of multiple indirections and control flow statements. The kernels used cover a variety of different access patterns, namely pointer traversals, indirection arrays, dynamic loop bounds and run-time dependent if-conditions. The kernels are small enough to be fully understood which makes this benchmark set very suitable for the evaluation of restructuring transformations.

## 1 Introduction

Optimizing compilers play a important role in the overall performance of applications. Restructuring transformations such as described for instance in [?] targeting the order of execution have proved very successful. However, these transformations all target regular code. In other words, loop bounds are known (at least symbolically) at compile-time and the associated iteration space can be described by linear inequalities. If all index functions are defined by affine transformations then full dependency information can be determined and any restructuring transformation preserving these dependencies can be applied.

In the other case, if the iteration space cannot directly be defined by a system of linear inequalities, the index functions use data that is only available at run-time or pointer traversals are encountered. Then the transformations as mentioned above cannot be applied. Nevertheless, these types of constructs regularly occur in today's applications. Thefore it is not suprising that a considerable research effort has been spent on improving the compiler effectiveness on irregular constructs, such as: conceptualizing specific data structures [?, ?], applying structure splitting, field reordering, array regrouping [?, ?, ?, ?], special prefetching techniques [?, ?, ?], symbolic compiler analysis [?] and providing run-time libraries [?, ?].

Therefore, research in optimizing compilers should focus on irregular applications. However, up till now, no evaluation platform exists that is designed specifically for the evaluation of optimizing compilers that target codes with a high degree of irregularity. In this paper, we propose a set of benchmarks that explicitly targets the evaluation of optimizing compilers for irregular codes. The benchmark suite consists of two subsets, one which uses pointer traversals and one in which the irregularity stems from the use of indirection arrays.

Many different benchmark suites have been implemented. Most of these benchmarks either target whole applications or regular kernels. Well known are the SPEC benchmarks [?] (e.g. SPEC CPU2000 and SPEC CPU2006). These benchmarks are taken from real applications and therefore are certainly useful for the assessment of a computer system as a whole. However, these applications are considerably larger than our kernels and therefore it is more difficult to evaluate and understand the effect of compiler transformation techniques. The only benchmark we used from SPEC CPU2000 is MCF, as its pointer traversal patterns are relatively simple. The Dhrystone benchmark [?, ?] is a synthetic benchmark used to measure CPU performance. It does not address highly irregular codes and therefore it is not suitable in our context. The same holds for Whetstone [?, ?], a floating point benchmark. A few benchmarks addressing irregular codes exist. Spark98 [?] (not to be confused with SPARK [?] and SPARK00) is a set of sparse matrix kernels for shared memory and message passing systems. It implements 5 programs, each of which performs some matrix multiplications. SPARK00 does not focus on matrix multiplication specifically, although matrix multiplication is part of our benchmark suite. Contrary to Spark98, we do not address parallel implementations of our benchmarks. Of course, if the compiler under evaluation does perform automatic parallelization, this is a perfectly legal code transformation. The SPARK [?] benchmark is a benchmark written in FORTRAN which aimed to analyze the interaction between the machine and the algorithm. We have translated some of SPARK's benchmark to C, which constitute the array-based group of benchmarks in SPARK00. However, SPARK did not contain pointer-based codes, which surely constitute a significant part of today's applications.

This paper is organized as follows. In Section 2, a description of all benchmarks is given. Section 3 describes the input data sets used in SPARK00. Irregular applications contain specific memory access patterns. Section 4 describes common patterns encountered in irregular applications. In Section 5, the benchmark framework is presented and it is shown how to process the results obtained by running the benchmarks. As an illustration, we present a small case study of the GCC compiler running on the Intel Itanium platform in Section 6. In Section 7, we summarize the work presented here and discuss our findings.

## 2 The Benchmarks

The two main sources causing irregularity are pointer traversals and the use of indirection arrays. We will first consider the pointer traversals in more detail. A pointer causes irregular access because its value can often not be determined at compile-time, especially in the case when the pointer is pointing to dynamically allocated memory, which is the case in nearly all applications that build dynamic data structures. Recursive data structures that are traversed and whose data members are accessed cause unpredictable access patterns and cannot be handled by regular transformation techniques. As mentioned in the introduction, SPARK00 consists of two subsets of benchmarks, one targeting pointer-based applications and the second, which is based on the SPARK [?] benchmark, targeting array-based applications. For the first subset, we used the SPARSE [?] library, with which we have implemented some direct and iterative methods for sparse matrices. These benchmarks have a varying degree of complexity, both in the complexity of the code as well as in the number of levels of indirection. A benchmark with even more complex access patterns is also included, namely MCF [?], which solves the Minimum Cost Flow problem. MCF is a program from the SPEC CPU2000 benchmark suite [?] and as such it is not included in the SPARK00 distribution. When users also want to include MCF in their experiments, they are adviced to get a separate SPEC CPU2000 license and obtain MCF directly from SPEC [?]. For each benchmark, we describe its structure and access pattern structure, which is dependent on the input data.

1. *SPMATVEC. Sparse matrix times dense vector.* The sparse matrix is represented using compressed row storage. The rows themselves are stored using linked lists. Each row is traversed and each element is multiplied with the corresponding element in the dense vector. The pointer traversal is one cause for irregularity. The other cause of irregularity is the indexing of the dense vector by a structure member of the linked list nodes (the column index of the sparse matrix element is used to index the vector). The result is stored in a separate dense vector.

2. *SPMATMAT. Sparse matrix times dense matrix.* The sparse matrix is represented using compressed row storage, the same manner as in SPMATVEC. The dense matrix is a C-style 2-dimensional array, which is dynamically allocated. This is different from FORTRAN-style 2-dimensional arrays, in which a contiguous block is accessed by an affine function of the loop index variables. Therefore, to access an element, two indirections are needed to access the appropriate value in C. The main difference with SPMATVEC is that in this benchmark, values are indirectly accessed whereas in the other benchmark, pointers are indirectly accessed.

3. *JACIT. Jacobi iteration.* Jacobi iteration [?] is used to solve $Ax = b$. The sparse matrix $A$ is represented using linked lists, compressed row storage. The linked list is traversed using two subsequent loops. One loop handles the elements before the diagonal and the second handles the elements after the diagonal. This traversal which is spread over two *while*-loops involves a termination condition in the first *while*-loop which is input data dependent.

4. *DSOLVE. Solve a linear system $Ax = b$ using forward substitution and backward substitution.* The matrix is represented using orthogonal linked lists (the matrix is traversed both row-wise and column-wise). The procedure takes a matrix that has been LU-factorized and solves $Ax = b$. In order to do this, the right hand side vector is permuted into an intermediate vector, after which forward substitution is applied to solve $Lc = b$. The forward substitution traverses the matrix column-wise. Next, $Uc = x$ can be solved by backward substitution which traverses the matrix row-wise. Finally, $x$ is permuted to obtain the result in the desired order.

5. *PCG. Preconditioned conjugate gradient.* PCG iteratively solves $Ax = b$. It uses the compressed row storage scheme implemented using linked lists. The code features indirect access caused by pointer traversals, as well as an array indexed by structure members of the list nodes. This array is also used in a regular fashion. Although the main computational part of PCG is the same as SPMATVEC, PCG uses the outcome of the multiplication in subsequent dot product operations.

6. *MCF. Minimum cost flow problem solver.* 181.mcf [?] is a program from the SPEC CPU2000 [?] benchmark suite that solves the minimum cost flow problem. The network simplex implementation is a pointer intensive application that is known to exhibit very poor cache performance due to the irregular nature of the memory access patterns caused by extensive used of pointer-linked data structures. Note that if this kernel is to be run, it should be separately licensed from SPEC.

The second subset consists of codes in which the irregular access originates from the use of indirection arrays. These codes are from the SPARK benchmark suite and have been translated to C.

1. *ASM. Assemble stiffness matrix.* Finite element methods involve an assembly step, in which all interactions between sub-elements consisting of 3-node triangular element are merged into one global matrix. Access to this matrix is governed by the connectivity matrix which is used to index the global array. The input data set used for this benchmark is the *wrench* data set, which is depicted in Figure 1.

2. *TRMAT. Transpose a matrix.* Computing the transpose of a sparse matrix contains quite some irregularity. First of all, the number of elements in a column is not known beforehand, and a traversal of the old index structure is needed to accumulate the right number of elements per column. This results in many scattered updates. The column counts are then translated into array offsets, which is done by a regular loop with one read-after-write dependency. Next, all data and index elements from the original matrix are traversed and mapped to the corresponding locations in the new arrays, causing single and double indirect access to arrays. The vector containing the row offsets is used to remember the current row offset within the target matrix. As a result, all elements of this vector must be moved one position to the right after filling the column and data vectors.

3. *CMcK. Compute Cuthill-McKee ordering.* The Cuthill-McKee method [?] computes a permutation array that aims to reduce the bandwidth of a sparse matrix. It does so by interpreting the sparse matrix as an adjacency matrix and computes a relabeling of the nodes. The relabeling is computed as follows. A breadth-first search is started at the node within minimal degree, which is labeled 1. Next, all adjacent nodes are considered and relabeled, starting with the node with lowest degree. The relabeling is recorded in the permutation array. The newly labeled nodes are expanded (following the ordering defined by the new labeling) and all unlabeled nodes are relabeled. This process continues until the entire connected component to which the starting node belongs is relabeled. If there still are nodes left, a remaining node with minimum degree is picked and the process above is repeated, until all nodes have been relabeled. The irregularity stems from the permutation array and the array that stores the column indices. The permutation array is used to locate the nodes that must be traversed next during the breadth-first search. Loop bounds and conditional branches are data dependent, which further complicates analysis.

4. *MPERM. Perform a symmetric permutation* $B = PAP^T$ *of an array $A$ and its associated right hand side vector $b$.* where $P$ is the permutation matrix. Instead of storing the permutation matrix, the mapping is stored in an array. Irregularity occurs naturally in permutation problems. The permutation requires a complete scan of the row index array to determine the new row sizes. This traversal mixes both regular and irregular access. Using the new row sizes, the new offsets are computed. Next, the iteration space of the newly generated index structure is traversed and the corresponding data from the original data structure is copied, which involves indirect accesses.

## 3 The Input Data

As input data sets for the current release of SPARK00, the following matrices have been selected: *add32, utm5940, sherman3, codecs4812.dc* and *bcsstk13*. All these matrices, with the exception of *codecs4812.dc*, are taken from the Harwell-Boeing Matrix Collection [?]. The matrix *codecs4812.dc* is part of the distribution of the SPARSE library [?]. Table 1 gives the main characteristics of the matrices and Figure 1 shows an overview of the structure of the matrices. Each structure plot shows a small region that is magnified, to show the diversity of the non-zero structures, which is not visible in the overview. Real world problems are often described by matrices where most elements are relatively close to the main diagonal. The matrices used reflect this fact. At first sight, the matrices might all look symmetric. However, only *bcsstk13* is symmetric. *add32* and *sherman3* are structurally symmetric (but the values are not symmetric) and

| Matrix | **add32** | **utm5940** | **sherman3** | **codecs4812.dc** | **bcsstk13** |
|---|---|---|---|---|---|
| Size | $4960 \times 4960$ | $5940 \times 5940$ | $5005 \times 5005$ | $4812 \times 4812$ | $2003 \times 2003$ |
| Entries | 23884 | 83842 | 20033 | 45192 | 42943 |
| Symmetric | No | No | Structural | No | Yes |

Table 1: Matrix characteristics

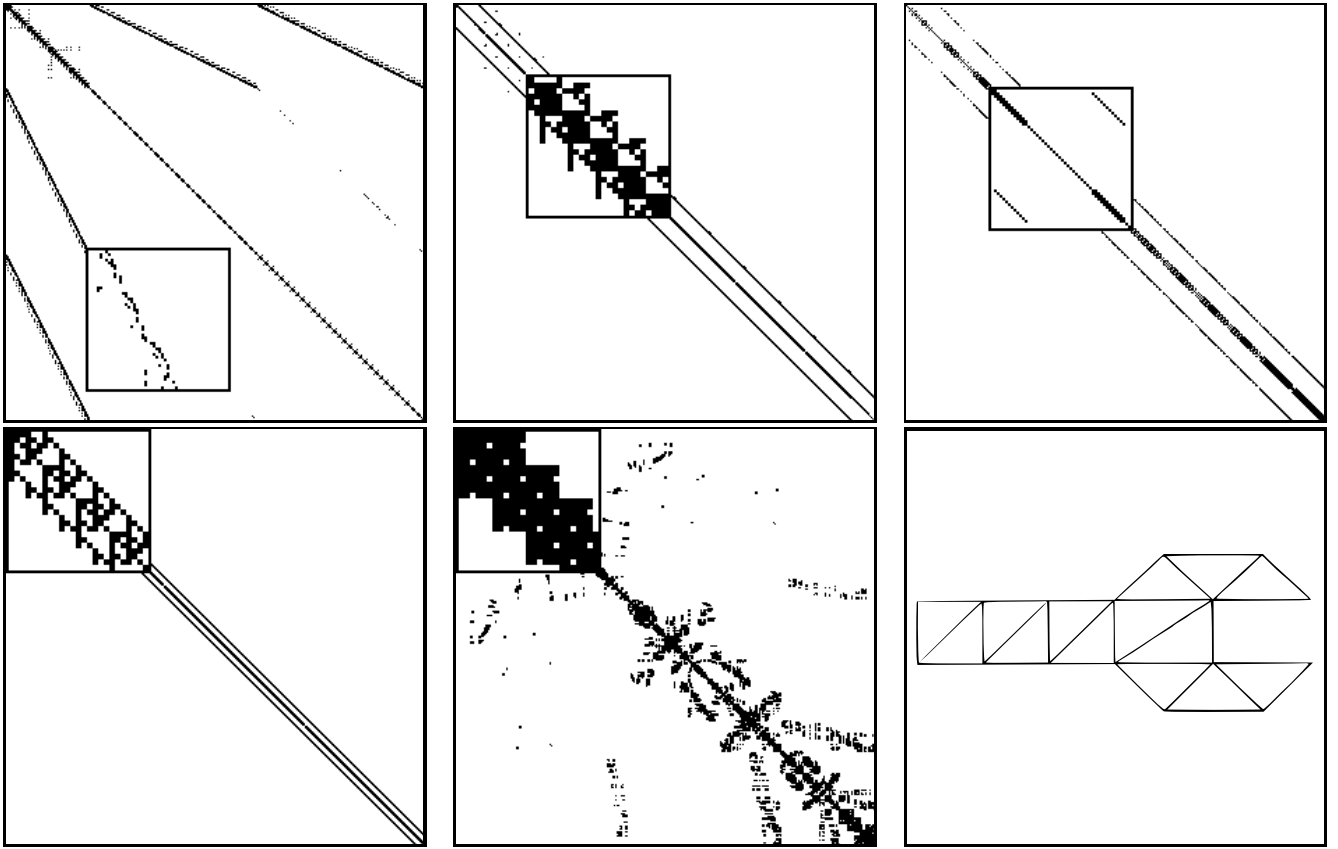

Figure 1: Input sets used in SPARK00. In clockwise order: add32, utm5940, sherman3, wrench, bcsstk13 and codecs4812.dc

*utm5940* and *codecs4812.dc* are unsymmetric. As for the benchmark CMcK symmetric matrices are required, the unsymmetric matrices are converted to symmetric matrices by mirroring the lower triangle of the matrix.

The selection of these matrices is based on the following criteria: diversity in application domain, variety in non-zero density and structure. *add32* is characterized by a dense diagonal, together with some additional elements which do lie on a specific band. These bands are however not parallel to the diagonal and therefore this matrix is significantly different in structure from the other matrices. *utm5940* is a matrix whose elements are mostly found on relatively dense diagonals and with some elements scattered in the upper left region. The structure of the diagonals is not symmetric. *sherman3* is a matrix whose elements are stored in diagonals. The diagonals themselves are thin, but very dense. *bcsstk13* is a matrix with a dense diagonal, and many off-diagonal clusters.

MCF uses the test data set from the SPEC 2000 benchmark suite. The ASM kernel uses the *wrench* data input set as described in [**?**]. For MPERM, the permutation matrix which results from the CMcK benchmark is taken and applied on the corresponding matrix. As stated above, for CMcK, symmetric matrices are generated using the lower triangle of the input matrix.

## 4 Irregular Code Characteristics

Irregular computations involve the execution of overhead code that is related to the navigation through data structures. As opposed to regular computations, the access patterns of such codes are inherently unpredictable and on the global application level, understanding their interaction is hard, maybe even impossible for a compiler.

In this section, we will describe some regularly occurring code characteristics in irregular applications. These characteristics are illustrated with examples drawn from the SPARK00 benchmark suite.

**Pointer chasing** Pointer-based applications involve the traversal of recursive pointer structures. These pointer structures inherently cause irregular accesses. Usually after fetching the next pointer, the memory pointed to is accessed. This pattern is found in all pointer-based benchmarks of SPARK00. The algorithms operating on sparse matrix structures use linked lists for the matrix representation. MCF implements a network structure, which is represented using pointer links. Pointer chasing prevents any restructuring transformation, as the access patterns are essentially serialized by the pointer structure. Moreover, the termination condition when traversing pointers in a loop is often dependent on run-time data, resulting in even more irregular behavior. See for example the following code which is taken from JACIT:

```
pElement = Matrix->FirstInRow[i];
while( pElement && pElement->Col < i ) {
  x_2[i] -= pElement->Real *
           x_1[ pElement->Col ];
  pElement = pElement->NextInRow;
}
```

**Inner versus outer pointer traversal** Pointer traversals do have a significant impact on program performance. However, it does make a difference if this traversal is an innermost traversal. In this case, the performance penalty is the biggest. Outer traversals have less impact on the performance and are susceptible to prefetching techniques, if there is a significant amount of work to be done before the next node in the pointer structure is accessed. Examples of kernels that contain inner traversals are: SPMATVEC, JACIT, DSOLVE, PCG and MCF. SPMATMAT contains an outer traversal. The following example taken from SPMATVEC shows an inner traversal:

```
for( row = 1; row <= left->Size; row++ ) {
  result[row] = 0.0f;
  pElement = left->FirstInRow[row];
  while( pElement ) {
    result[row] += pElement->Real *
                   right[pElement->Col];
    pElement = pElement->NextInRow;
  }
}
```

The outer traversal in SPMATMAT looks as follows:

```
for( row = 1; row <= left->Size; row++ ) {
  pElement = left->FirstInRow[row];
  while( pElement ) {
    for( col = 1; col <= cols; col++ ) {
      result[row][col] += pElement->Real *
      right[pElement->Col][col];
    }
    pElement = pElement->NextInRow;
  }
}
```

**Array indirection by structure member** If pointer-based and array-based code is mixed, structure members can be used to index arrays. All pointer-based codes from SPARK00 contain this pattern. For example, this statement is from MCF:

```
arc2 = arc->head[net->n_trips].firstout;
```

This pattern also appears in the examples given above.

**Array indirection by indirection array** If all code is array-based, indirection is done using indirection arrays. This pattern is used extensively in array-based sparse matrix codes, which store their values and indexing information in separate arrays. All array-based codes from SPARK00 contain this pattern. The following example is from MPERM. The loops fill the new data and index arrays.

```
for( ii = 1; ii <= n; ii++ ) {
  k0 = ia[ iord[ii] ] - iao[ii];
  for( k = iao[ii];
       k <= iao[ii+1]-1; k++ ) {
    jao[k] = riord[ ja[k0+k] ];
    ao[k] = a[k0+k];
  }
}
```

**Dynamic loop bounds** One of the reasons that prevents dependency analysis is the occurrence of loop bounds that cannot be determined at compile-time. This pattern naturally occurs in permutation problems, where indirection is used to determine a particular region needed for computation. An example of this pattern is found in CMcK:

```
for(k = pntr[parray[j]];
        k <= pntr[parray[j]+1]-1; k++) {
  tmp = rwind[k];
  if( ! consd[tmp] ) continue;
  deg = pntr[tmp+1] - pntr[tmp] - 1;
  if( deg >= low ) continue;
  low = deg;
  lowl = tmp;
}
```

This code traverses a row of a matrix which has been determined at previous stage of the algorithm and stored in *parray*. MPERM and TRMAT contain similar constructs.

**Gather-scatter pattern** The gather-scatter pattern is a pattern that is encountered often in irregular codes. The gather operation is performed to bring data into a suitable form for further manipulation. For instance, in DSOLVE, the input matrix is factorized and thus reordered, but the programming interface expects dense right hand side arrays as input in the original order. The gather operation is performed to permute the input array. Upon completion of the routine, the result is written back to the solution array in the original order.

```
/* Gather */
pExtOrder = &Matrix->IntToExtRowMap[Size];
for (I = Size; I > 0; I--)
  Intermediate[I] = RHS[*(pExtOrder--)];

/* Scatter */
pExtOrder = &Matrix->IntToExtColMap[Size];
for (I = Size; I > 0; I--)
  Solution[*(pExtOrder--)] = Intermediate[I];
```

# 5 Running SPARK00

In this section, we describe how the SPARK00 benchmark suite should be used. At the top level directory, the SPARK00 package has the following structure:

```
spark00/
    bench/
    bin/
    data/
    exp/
    include/
    lib/
    results/
```

The *bench* directory contains all the benchmarks. Each benchmark contains a *run* directory, with a file *Run.pl*. This file defines how a specific benchmark is executed. The *bin* directory contains *runspark.pl*, which is the script that executes all benchmarks. The *data* directory contains all input data sets used by the benchmarks. Global configuration and the benchmark execution framework can be found in the *lib* directory. Execution times are written to separate files within a directory under *results*, which carries the name of the benchmark configuration as specified by the user. The name *base* should be used for the baseline configuration. The final results are aggregated into one file, which is written to the directory *exp/data*.

Configuration of SPARK00 is done by copying the default configuration file *vars.mk.def* in the directory *lib/mk* to *vars.mk*. Within *vars.mk*, the variables *CC* and *CFLAGS* are used to configure a benchmark run. *CC* defines the compiler executable used while *CFLAGS* is used to pass parameters to the compiler. SPARK00 is executed by running *perl -Mlib=../lib runspark.pl* from the *bin* directory.

The final results can be found in the file *spark.dat* in the directory *exp/data*. Each entry in this file has the following format: *id benchmark matrix reftime time*. *id* denotes the configuration used, benchmark is the name of the benchmark, and matrix is the input matrix used. In the case of MCF and ASM, the matrix entry is set to *none*. Both the reference time and the time measured for this specific configuration are stored in the columns *reftime* and *time*, respectively.

# 6  Case Study: GCC

In this section we present a case study using the GCC 4.2.2 compiler, which is the most recent version from the release series. The experiments have been conducted on an Intel Itanium 2 platform, running Red Hat Enterprise Linux AS release 3. The Itanium architecture relies heavily on explicit parallelism, that is, the compiler is largely responsible for identifying parallelism and instruction scheduling. Optimizing applications for such architectures in the presence of irregular access patterns is therefore very challenging. In this experiment, two different configurations are compared to a reference benchmark, which is GCC 4.2.2 without any optimizations. The two configurations are *-O2* and *-O3*. Figure 2 shows the result of running SPARK00 on the Itanium architecture. Each plot depicts the result of all the benchmarks for a specific matrix. MCF and ASM are shown is a separate plot, as these two have different input data sets.

Most striking is the difference in speedup between pointer-based applications (SPMATVEC, SPMATMAT, JACIT, DSOLVE, PCG and MCF) versus indirection array applications (ASM, TRMAT, CMcK and MPERM). Both type of applications exhibit a high degree of irregularity, but the clustered storage as found in the array-based codes appear to offer better optimization opportunities. An explanation for this difference is that some operations on the arrays can be done sequentially, whereas such an operation on pointer-based codes still involves the traversal of a pointer-linked structure, causing an indirection at every node of the data structure.

Viewing the results in the context of the characteristics defined in Section 4, reveals some interesting points. Pointer chasing clearly poses a problem for optimizing compilers, as we can see that the results on optimizing the pointer-based benchmarks shows less improvement than optimizations on array-based codes. The outer traversal which is found in SPMATMAT has less influence on the optimization results, as expected. For data sets with relatively long linked lists, such as *bcsstk13*, this puts a higher computational burden on the inner traversal, which has a huge impact on the performance. In addition, the inner loop bodies of the pointer-based benchmarks use arrays indexed by structure members, which also results in scattered memory access. As a result, all benchmarks containing an inner traversal that are used heavily do barely benefit from the compiler optimizations. The same applies to MCF, whose extensive use of pointer traversals prevents proper optimization.

If we consider the differences between the various compiler settings, it becomes clear that the optimizations are clearly effective in some cases. However, in general there is not a large difference between the different compiler settings.

# 7  Conclusions

In this paper, we have outlined SPARK00, a benchmark suite that specifically targets irregular applications to be used for the evaluation of optimizing compilers. Contrary to other benchmarks, we isolated the irregularity of applications in computational kernels. This allows for evaluating the effectiveness of compiler transformations and relating performance to specific irregularity characteristics.

As an illustration of the usage of SPARK00, we have run the benchmarks using GCC 4.2.2 on the Intel Itanium architecture. In this paper, we did not intend to provide a complete analysis of the effectiveness of GCC 4.2.2, but rather we showed what can be done if SPARK00 is being used. In forthcoming publications, we will perform an extensive analysis of compiler transformations across different platforms.

We realize that our benchmarks do not completely represent the behavior of full applications. However, understanding the implications of compiler optimizations on the full application level is a difficult task and as long as state-of-the-art compilers fail to handle irregular applications effectively, our benchmark suite will provide a effective way of evaluating compiler transformation techniques.

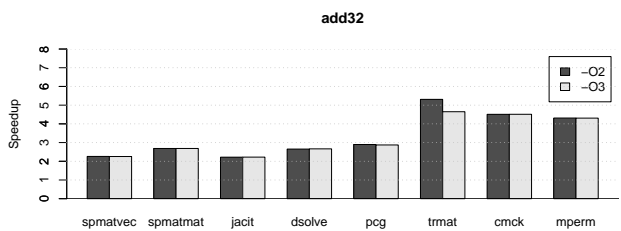
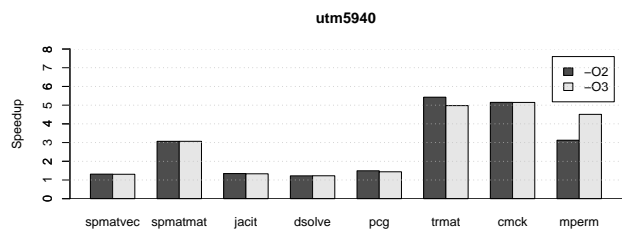
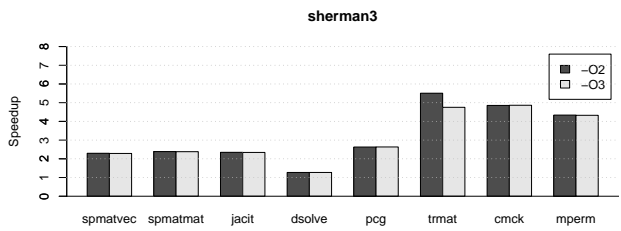
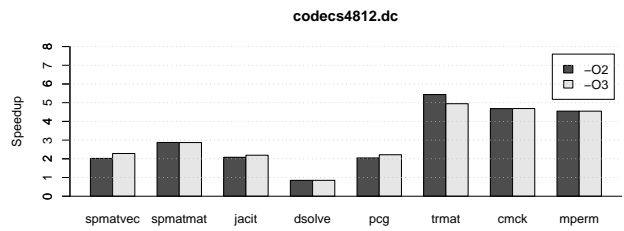
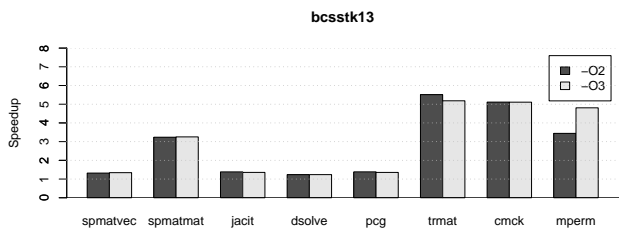
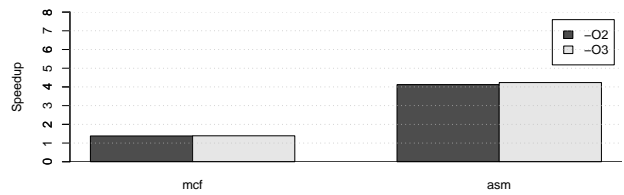

Figure 2: SPARK00 results for GCC on the Intel Itanium 2 architecture

# SPARK00: A Benchmark Package
# for the Compiler Evaluation of Irregular/Sparse Codes

H.L.A. van der Spek
E.M. Bakker
H.A.G. Wijshoff
Leiden University, The Netherlands

**Abstract**

We propose a set of benchmarks that specifically targets a major cause of performance degradation in high performance computing platforms: irregular access patterns. These benchmarks are meant to be used to asses the performance of optimizing compilers on codes with a varying degree of irregular access. The irregularity caused by the use of pointers and indirection arrays are a major challenge for optimizing compilers. Codes containing such patterns are notoriously hard to optimize but they have a huge impact on the performance of modern architectures, which are under-utilized when encountering irregular memory accesses. In this paper, a set of benchmarks is described that explicitly measures the performance of kernels containing a variety of different access patterns found in real world applications. By offering a varying degree of complexity, we provide a platform for measuring the effectiveness of transformations. The difference in complexity stems from a difference in traversal patterns, the use of multiple indirections and control flow statements. The kernels used cover a variety of different access patterns, namely pointer traversals, indirection arrays, dynamic loop bounds and run-time dependent if-conditions. The kernels are small enough to be fully understood which makes this benchmark set very suitable for the evaluation of restructuring transformations.

## 1 Introduction

Optimizing compilers play a important role in the overall performance of applications. Restructuring transformations such as described for instance in [16] targeting the order of execution have proved very successful. However, these transformations all target regular code. In other words, loop bounds are known (at least symbolically) at compile-time and the associated iteration space can be described by linear inequalities. If all index functions are defined by affine transformations then full dependency information can be determined and any restructuring transformation preserving these dependencies can be applied.

In the other case, if the iteration space cannot directly be defined by a system of linear inequalities, the index functions use data that is only available at run-time or pointer traversals are encountered. Then the transformations as mentioned above cannot be applied. Nevertheless, these types of constructs regularly occur in today's applications. Thefore it is not suprising that a considerable research effort has been spent on improving the compiler effectiveness on irregular constructs, such as: conceptualizing specific data structures [1, 18], applying structure splitting, field reordering, array regrouping [2, 27, 8, 6], special prefetching techniques [10, 14, 26], symbolic compiler analysis [12] and providing run-time libraries [19, 21].

Therefore, research in optimizing compilers should focus on irregular applications. However, up till now, no evaluation platform exists that is designed specifically for the evaluation of optimizing compilers that target codes with a high degree of irregularity. In this paper, we propose a set of benchmarks that explicitly targets the evaluation of optimizing compilers for irregular codes. The benchmark suite consists of two subsets, one which uses pointer traversals and one in which the irregularity stems from the use of indirection arrays.

Many different benchmark suites have been implemented. Most of these benchmarks either target whole applications or regular kernels. Well known are the SPEC benchmarks [23] (e.g. SPEC CPU2000 and SPEC CPU2006). These benchmarks are taken from real applications and therefore are certainly useful for the assessment of a computer system as a whole. However, these applications are considerably larger than our kernels and therefore it is more difficult to evaluate and understand the effect of compiler transformation techniques. The only benchmark we used from SPEC CPU2000 is MCF, as its pointer traversal patterns are relatively simple. The Dhrystone benchmark [25, 24] is a synthetic benchmark used to measure CPU performance. It does not address highly irregular codes and therefore it is not suitable in our context. The same holds for Whetstone [3, 17], a floating point benchmark. A few benchmarks addressing irregular codes exist. Spark98 [15] (not to be confused with SPARK [20] and SPARK00) is a set of sparse matrix kernels for shared memory and message passing systems. It implements 5 programs, each of which performs some matrix multiplications. SPARK00 does not focus on matrix multiplication specifically, although matrix multiplication is part of our benchmark suite. Contrary to Spark98, we do not address parallel implementations of our benchmarks. Of course, if the compiler under evaluation does perform automatic parallelization, this is a perfectly legal code transformation. The SPARK [20] benchmark is a benchmark written in FORTRAN which aimed to analyze the interaction between the machine and the algorithm. We have translated some of SPARK's benchmark to C, which constitute the array-based group of benchmarks in SPARK00. However, SPARK did not contain pointer-based codes, which surely constitute a significant part of today's applications.

This paper is organized as follows. In Section 2, a description of all benchmarks is given. Section 3 describes the input data sets used in SPARK00. Irregular applications contain specific memory access patterns. Section 4 describes common patterns encountered in irregular applications. In Section 5, the benchmark framework is presented and it is shown how to process the results obtained by running the benchmarks. As an illustration, we present a small case study of the GCC compiler running on the Intel Itanium platform in Section 6. In Section 7, we summarize the work presented here and discuss our findings.

## 2 The Benchmarks

The two main sources causing irregularity are pointer traversals and the use of indirection arrays. We will first consider the pointer traversals in more detail. A pointer causes irregular access because its value can often not be determined at compile-time, especially in the case when the pointer is pointing to dynamically allocated memory, which is the case in nearly all applications that build dynamic data structures. Recursive data structures that are traversed and whose data members are accessed cause unpredictable access patterns and cannot be handled by regular transformation techniques. As mentioned in the introduction, SPARK00 consists of two subsets of benchmarks, one targeting pointer-based applications and the second, which is based on the SPARK [20] benchmark, targeting array-based applications. For the first subset, we used the SPARSE [11] library, with which we have implemented some direct and iterative methods for sparse matrices. These benchmarks have a varying degree of complexity, both in the complexity of the code as well as in the number of levels of indirection. A benchmark with even more complex access patterns is also included, namely MCF [13], which solves the Minimum Cost Flow problem. MCF is a program from the SPEC CPU2000 benchmark suite [9] and as such it is not included in the SPARK00 distribution. When users also want to include MCF in their experiments, they are advised to get a separate SPEC CPU2000 license and obtain MCF directly from SPEC [23]. For each benchmark, we describe its structure and access pattern structure, which is dependent on the input data.

1. *SPMATVEC. Sparse matrix times dense vector.* The sparse matrix is represented using compressed row storage. The rows themselves are stored using linked lists. Each row is traversed and each element is multiplied with the corresponding element in the dense vector. The pointer traversal is one cause for irregularity. The other cause of irregularity is the indexing of the dense vector by a structure member of the linked list nodes (the column index of the sparse matrix element is used to index the vector). The result is stored in a separate dense vector.

2. *SPMATMAT. Sparse matrix times dense matrix.* The sparse matrix is represented using compressed row storage, the same manner as in SPMATVEC. The dense matrix is a C-style 2-dimensional array, which is dynamically allocated. This is different from FORTRAN-style 2-dimensional arrays, in which a contiguous block is accessed by an affine function of the loop index variables. Therefore, to access an element, two indirections are needed to access the appropriate value in C. The main difference with SPMATVEC is that in this benchmark, values are indirectly accessed whereas in the other benchmark, pointers are indirectly accessed.

3. *JACIT. Jacobi iteration.* Jacobi iteration [7] is used to solve $Ax = b$. The sparse matrix $A$ is represented using linked lists, compressed row storage. The linked list is traversed using two subsequent loops. One loop handles the elements before the diagonal and the second handles the elements after the diagonal. This traversal which is spread over two *while*-loops involves a termination condition in the first *while*-loop which is input data dependent.

4. *DSOLVE. Solve a linear system $Ax = b$ using forward substitution and backward substitution.* The matrix is represented using orthogonal linked lists (the matrix is traversed both row-wise and column-wise). The procedure takes a matrix that has been LU-factorized and solves $Ax = b$. In order to do this, the right hand side vector is permuted into an intermediate vector, after which forward substitution is applied to solve $Lc = b$. The forward substitution traverses the matrix column-wise. Next, $Uc = x$ can be solved by backward substitution which traverses the matrix row-wise. Finally, $x$ is permuted to obtain the result in the desired order.

5. *PCG. Preconditioned conjugate gradient.* PCG iteratively solves $Ax = b$. It uses the compressed row storage scheme implemented using linked lists. The code features indirect access caused by pointer traversals, as well as an array indexed by structure members of the list nodes. This array is also used in a regular fashion. Although the main computational part of PCG is the same as SPMATVEC, PCG uses the outcome of the multiplication in subsequent dot product operations.

6. *MCF. Minimum cost flow problem solver.* 181.mcf [13] is a program from the SPEC CPU2000 [9] benchmark suite that solves the minimum cost flow problem. The network simplex implementation is a pointer intensive application that is known to exhibit very poor cache performance due to the irregular nature of the memory access patterns caused by extensive used of pointer-linked data structures. Note that if this kernel is to be run, it should be separately licensed from SPEC.

The second subset consists of codes in which the irregular access originates from the use of indirection arrays. These codes are from the SPARK benchmark suite and have been translated to C.

1. *ASM. Assemble stiffness matrix.* Finite element methods involve an assembly step, in which all interactions between sub-elements consisting of 3-node triangular element are merged into one global matrix. Access to this matrix is governed by the connectivity matrix which is used to index the global array. The input data set used for this benchmark is the *wrench* data set, which is depicted in Figure 1.

2. *TRMAT. Transpose a matrix.* Computing the transpose of a sparse matrix contains quite some irregularity. First of all, the number of elements in a column is not known beforehand, and a traversal of the old index structure is needed to accumulate the right number of elements per column. This results in many scattered updates. The column counts are then translated into array offsets, which is done by a regular loop with one read-after-write dependency. Next, all data and index elements from the original matrix are traversed and mapped to the corresponding locations in the new arrays, causing single and double indirect access to arrays. The vector containing the row offsets is used to remember the current row offset within the target matrix. As a result, all elements of this vector must be moved one position to the right after filling the column and data vectors.

3. *CMcK. Compute Cuthill-McKee ordering.* The Cuthill-McKee method [4] computes a permutation array that aims to reduce the bandwidth of a sparse matrix. It does so by interpreting the sparse matrix as an adjacency matrix and computes a relabeling of the nodes. The relabeling is computed as follows. A breadth-first search is started at the node within minimal degree, which is labeled 1. Next, all adjacent nodes are considered and relabeled, starting with the node with lowest degree. The relabeling is recorded in the permutation array. The newly labeled nodes are expanded (following the ordering defined by the new labeling) and all unlabeled nodes are relabeled. This process continues until the entire connected component to which the starting node belongs is relabeled. If there still are nodes left, a remaining node with minimum degree is picked and the process above is repeated, until all nodes have been relabeled. The irregularity stems from the permutation array and the array that stores the column indices. The permutation array is used to locate the nodes that must be traversed next during the breadth-first search. Loop bounds and conditional branches are data dependent, which further complicates analysis.

4. *MPERM. Perform a symmetric permutation $B = PAP^T$ of an array $A$ and its associated right hand side vector $b$.* where $P$ is the permutation matrix. Instead of storing the permutation matrix, the mapping is stored in an array. Irregularity occurs naturally in permutation problems. The permutation requires a complete scan of the row index array to determine the new row sizes. This traversal mixes both regular and irregular access. Using the new row sizes, the new offsets are computed. Next, the iteration space of the newly generated index structure is traversed and the corresponding data from the original data structure is copied, which involves indirect accesses.

# 3 The Input Data

As input data sets for the current release of SPARK00, the following matrices have been selected: *add32, utm5940, sherman3, codecs4812.dc* and *bcsstk13*. All these matrices, with the exception of *codecs4812.dc*, are taken from the Harwell-Boeing Matrix Collection [5]. The matrix *codecs4812.dc* is part of the distribution of the SPARSE library [11]. Table 1 gives the main characteristics of the matrices and Figure 1 shows an overview of the structure of the matrices. Each structure plot shows a small region that is magnified, to show the diversity of the non-zero structures, which is not visible in the overview. Real world problems are often described by matrices where most elements are relatively close to the main diagonal. The matrices used reflect this fact. At first sight, the matrices might all look symmetric. However, only *bcsstk13* is symmetric. *add32* and *sherman3* are structurally symmetric (but the values are not symmetric) and

| Matrix | **add32** | **utm5940** | **sherman3** | **codecs4812.dc** | **bcsstk13** |
|---|---|---|---|---|---|
| Size | $4960 \times 4960$ | $5940 \times 5940$ | $5005 \times 5005$ | $4812 \times 4812$ | $2003 \times 2003$ |
| Entries | 23884 | 83842 | 20033 | 45192 | 42943 |
| Symmetric | No | No | Structural | No | Yes |

Table 1: Matrix characteristics

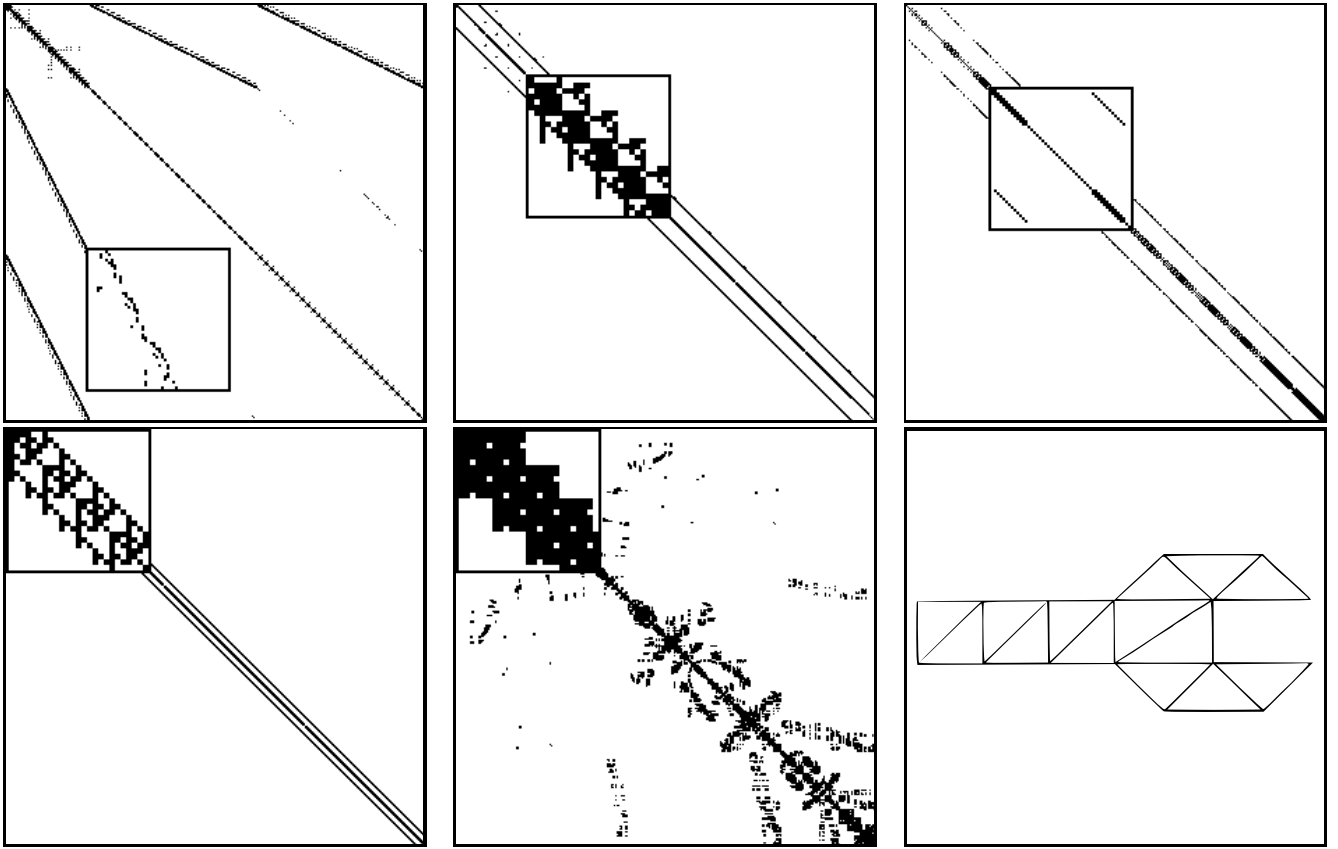

Figure 1: Input sets used in SPARK00. In clockwise order: add32, utm5940, sherman3, wrench, bcsstk13 and codecs4812.dc

*utm5940* and *codecs4812.dc* are unsymmetric. As for the benchmark CMcK symmetric matrices are required, the unsymmetric matrices are converted to symmetric matrices by mirroring the lower triangle of the matrix.

The selection of these matrices is based on the following criteria: diversity in application domain, variety in non-zero density and structure. *add32* is characterized by a dense diagonal, together with some additional elements which do lie on a specific band. These bands are however not parallel to the diagonal and therefore this matrix is significantly different in structure from the other matrices. *utm5940* is a matrix whose elements are mostly found on relatively dense diagonals and with some elements scattered in the upper left region. The structure of the diagonals is not symmetric. *sherman3* is a matrix whose elements are stored in diagonals. The diagonals themselves are thin, but very dense. *bcsstk13* is a matrix with a dense diagonal, and many off-diagonal clusters.

MCF uses the test data set from the SPEC 2000 benchmark suite. The ASM kernel uses the *wrench* data input set as described in [22]. For MPERM, the permutation matrix which results from the CMcK benchmark is taken and applied on the corresponding matrix. As stated above, for CMcK, symmetric matrices are generated using the lower triangle of the input matrix.

## 4 Irregular Code Characteristics

Irregular computations involve the execution of overhead code that is related to the navigation through data structures. As opposed to regular computations, the access patterns of such codes are inherently unpredictable and on the global application level, understanding their interaction is hard, maybe even impossible for a compiler.

In this section, we will describe some regularly occurring code characteristics in irregular applications. These characteristics are illustrated with examples drawn from the SPARK00 benchmark suite.

**Pointer chasing** Pointer-based applications involve the traversal of recursive pointer structures. These pointer structures inherently cause irregular accesses. Usually after fetching the next pointer, the memory pointed to is accessed. This pattern is found in all pointer-based benchmarks of SPARK00. The algorithms operating on sparse matrix structures use linked lists for the matrix representation. MCF implements a network structure, which is represented using pointer links. Pointer chasing prevents any restructuring transformation, as the access patterns are essentially serialized by the pointer structure. Moreover, the termination condition when traversing pointers in a loop is often dependent on run-time data, resulting in even more irregular behavior. See for example the following code which is taken from JACIT:

```
pElement = Matrix->FirstInRow[i];
while( pElement && pElement->Col < i ) {
  x_2[i] -= pElement->Real *
           x_1[ pElement->Col ];
  pElement = pElement->NextInRow;
}
```

**Inner versus outer pointer traversal** Pointer traversals do have a significant impact on program performance. However, it does make a difference if this traversal is an innermost traversal. In this case, the performance penalty is the biggest. Outer traversals have less impact on the performance and are susceptible to prefetching techniques, if there is a significant amount of work to be done before the next node in the pointer structure is accessed. Examples of kernels that contain inner traversals are: SPMATVEC, JACIT, DSOLVE, PCG and MCF. SPMATMAT contains an outer traversal. The following example taken from SPMATVEC shows an inner traversal:

```
for( row = 1; row <= left->Size; row++ ) {
  result[row] = 0.0f;
  pElement = left->FirstInRow[row];
  while( pElement ) {
    result[row] += pElement->Real *
                   right[pElement->Col];
    pElement = pElement->NextInRow;
  }
}
```

The outer traversal in SPMATMAT looks as follows:

```
for( row = 1; row <= left->Size; row++ ) {
  pElement = left->FirstInRow[row];
  while( pElement ) {
    for( col = 1; col <= cols; col++ ) {
      result[row][col] += pElement->Real *
      right[pElement->Col][col];
    }
    pElement = pElement->NextInRow;
  }
}
```

**Array indirection by structure member** If pointer-based and array-based code is mixed, structure members can be used to index arrays. All pointer-based codes from SPARK00 contain this pattern. For example, this statement is from MCF:

```
arc2 = arc->head[net->n_trips].firstout;
```

This pattern also appears in the examples given above.

**Array indirection by indirection array** If all code is array-based, indirection is done using indirection arrays. This pattern is used extensively in array-based sparse matrix codes, which store their values and indexing information in separate arrays. All array-based codes from SPARK00 contain this pattern. The following example is from MPERM. The loops fill the new data and index arrays.

```
for( ii = 1; ii <= n; ii++ ) {
  k0 = ia[ iord[ii] ] - iao[ii];
  for( k = iao[ii];
       k <= iao[ii+1]-1; k++ ) {
    jao[k] = riord[ ja[k0+k] ];
    ao[k] = a[k0+k];
  }
}
```

**Dynamic loop bounds** One of the reasons that prevents dependency analysis is the occurrence of loop bounds that cannot be determined at compile-time. This pattern naturally occurs in permutation problems, where indirection is used to determine a particular region needed for computation. An example of this pattern is found in CMcK:

```
for(k = pntr[parray[j]];
        k <= pntr[parray[j]+1]-1; k++) {
  tmp = rwind[k];
  if( ! consd[tmp] ) continue;
  deg = pntr[tmp+1] - pntr[tmp] - 1;
  if( deg >= low ) continue;
  low = deg;
  lowl = tmp;
}
```

This code traverses a row of a matrix which has been determined at previous stage of the algorithm and stored in *parray*. MPERM and TRMAT contain similar constructs.

**Gather-scatter pattern** The gather-scatter pattern is a pattern that is encountered often in irregular codes. The gather operation is performed to bring data into a suitable form for further manipulation. For instance, in DSOLVE, the input matrix is factorized and thus reordered, but the programming interface expects dense right hand side arrays as input in the original order. The gather operation is performed to permute the input array. Upon completion of the routine, the result is written back to the solution array in the original order.

```
/* Gather */
pExtOrder = &Matrix->IntToExtRowMap[Size];
for (I = Size; I > 0; I--)
  Intermediate[I] = RHS[*(pExtOrder--)];

/* Scatter */
pExtOrder = &Matrix->IntToExtColMap[Size];
for (I = Size; I > 0; I--)
  Solution[*(pExtOrder--)] = Intermediate[I];
```

# 5 Running SPARK00

In this section, we describe how the SPARK00 benchmark suite should be used. At the top level directory, the SPARK00 package has the following structure:

```
spark00/
    bench/
    bin/
    data/
    exp/
    include/
    lib/
    results/
```

The *bench* directory contains all the benchmarks. Each benchmark contains a *run* directory, with a file *Run.pl*. This file defines how a specific benchmark is executed. The *bin* directory contains *runspark.pl*, which is the script that executes all benchmarks. The *data* directory contains all input data sets used by the benchmarks. Global configuration and the benchmark execution framework can be found in the *lib* directory. Execution times are written to separate files within a directory under *results*, which carries the name of the benchmark configuration as specified by the user. The name *base* should be used for the baseline configuration. The final results are aggregated into one file, which is written to the directory *exp/data*.

Configuration of SPARK00 is done by copying the default configuration file *vars.mk.def* in the directory *lib/mk* to *vars.mk*. Within *vars.mk*, the variables *CC* and *CFLAGS* are used to configure a benchmark run. *CC* defines the compiler executable used while *CFLAGS* is used to pass parameters to the compiler. SPARK00 is executed by running *perl -Mlib=../lib runspark.pl* from the *bin* directory.

The final results can be found in the file *spark.dat* in the directory *exp/data*. Each entry in this file has the following format: *id benchmark matrix reftime time*. *id* denotes the configuration used, benchmark is the name of the benchmark, and matrix is the input matrix used. In the case of MCF and ASM, the matrix entry is set to *none*. Both the reference time and the time measured for this specific configuration are stored in the columns *reftime* and *time*, respectively.

# 6 Case Study: GCC

In this section we present a case study using the GCC 4.2.2 compiler, which is the most recent version from the release series. The experiments have been conducted on an Intel Itanium 2 platform, running Red Hat Enterprise Linux AS release 3. The Itanium architecture relies heavily on explicit parallelism, that is, the compiler is largely responsible for identifying parallelism and instruction scheduling. Optimizing applications for such architectures in the presence of irregular access patterns is therefore very challenging. In this experiment, two different configurations are compared to a reference benchmark, which is GCC 4.2.2 without any optimizations. The two configurations are *-O2* and *-O3*. Figure 2 shows the result of running SPARK00 on the Itanium architecture. Each plot depicts the result of all the benchmarks for a specific matrix. MCF and ASM are shown is a separate plot, as these two have different input data sets.

Most striking is the difference in speedup between pointer-based applications (SPMATVEC, SPMATMAT, JACIT, DSOLVE, PCG and MCF) versus indirection array applications (ASM, TRMAT, CMcK and MPERM). Both type of applications exhibit a high degree of irregularity, but the clustered storage as found in the array-based codes appear to offer better optimization opportunities. An explanation for this difference is that some operations on the arrays can be done sequentially, whereas such an operation on pointer-based codes still involves the traversal of a pointer-linked structure, causing an indirection at every node of the data structure.

Viewing the results in the context of the characteristics defined in Section 4, reveals some interesting points. Pointer chasing clearly poses a problem for optimizing compilers, as we can see that the results on optimizing the pointer-based benchmarks shows less improvement than optimizations on array-based codes. The outer traversal which is found in SPMATMAT has less influence on the optimization results, as expected. For data sets with relatively long linked lists, such as *bcsstk13*, this puts a higher computational burden on the inner traversal, which has a huge impact on the performance. In addition, the inner loop bodies of the pointer-based benchmarks use arrays indexed by structure members, which also results in scattered memory access. As a result, all benchmarks containing an inner traversal that are used heavily do barely benefit from the compiler optimizations. The same applies to MCF, whose extensive use of pointer traversals prevents proper optimization.

If we consider the differences between the various compiler settings, it becomes clear that the optimizations are clearly effective in some cases. However, in general there is not a large difference between the different compiler settings.

# 7 Conclusions

In this paper, we have outlined SPARK00, a benchmark suite that specifically targets irregular applications to be used for the evaluation of optimizing compilers. Contrary to other benchmarks, we isolated the irregularity of applications in computational kernels. This allows for evaluating the effectiveness of compiler transformations and relating performance to specific irregularity characteristics.

As an illustration of the usage of SPARK00, we have run the benchmarks using GCC 4.2.2 on the Intel Itanium architecture. In this paper, we did not intend to provide a complete analysis of the effectiveness of GCC 4.2.2, but rather we showed what can be done if SPARK00 is being used. In forthcoming publications, we will perform an extensive analysis of compiler transformations across different platforms.

We realize that our benchmarks do not completely represent the behavior of full applications. However, understanding the implications of compiler optimizations on the full application level is a difficult task and as long as state-of-the-art compilers fail to handle irregular applications effectively, our benchmark suite will provide a effective way of evaluating compiler transformation techniques.

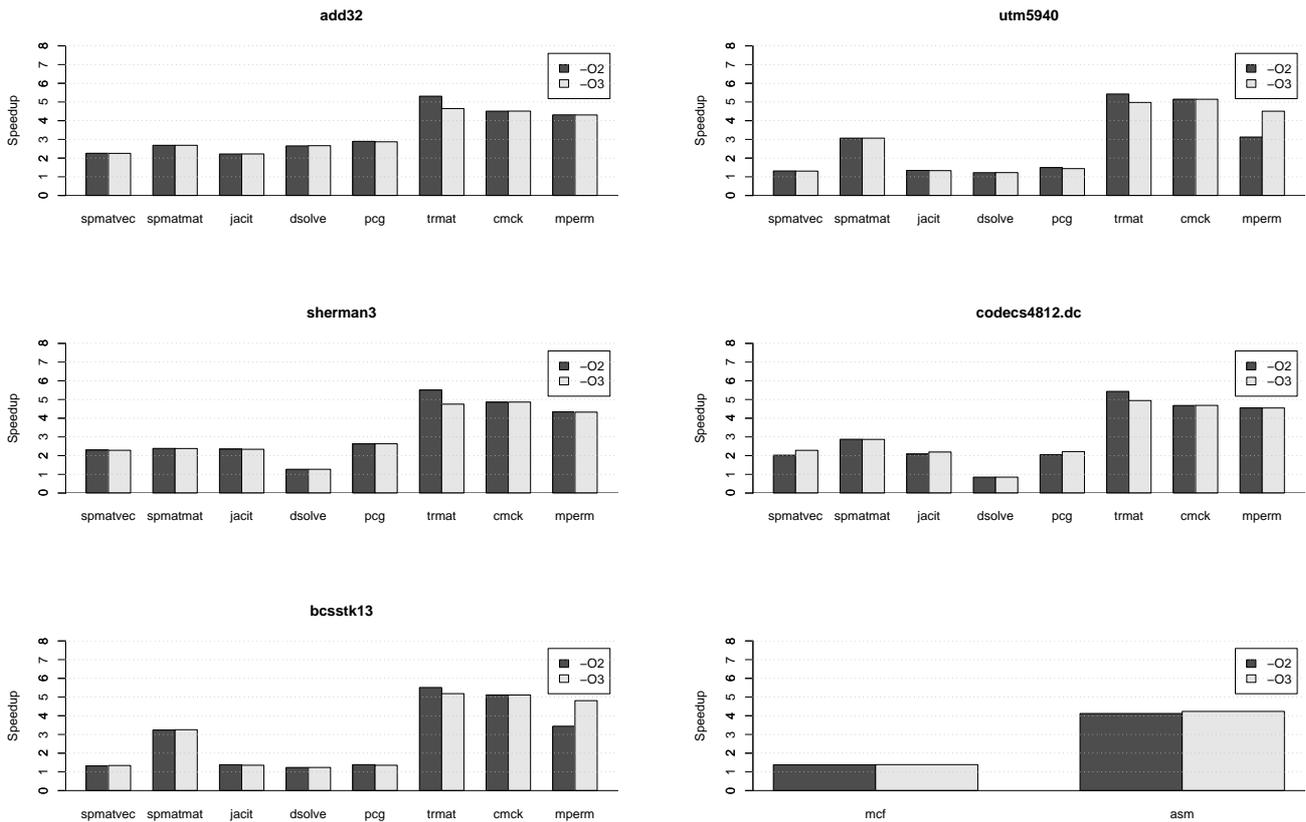

Figure 2: SPARK00 results for GCC on the Intel Itanium 2 architecture